\documentclass[12pt,twoside, a4paper]{article}
\def\pd{\partial}
\def\mc{\mathcal}

\usepackage[dvips]{graphicx}
\usepackage{amssymb}
\usepackage{amssymb,amsmath}
\usepackage{graphicx}
\input{epsf.sty}  
\begin{document}
\begin{center}
\LARGE{\textbf{Gaugings of four-dimensional $N=3$ supergravity and AdS$_4$/CFT$_3$ holography}}
\end{center}
\vspace{1 cm}
\begin{center}
\large{\textbf{Parinya Karndumri}$^a$ and \textbf{Khem
Upathambhakul}$^b$}
\end{center}
\begin{center}
String Theory and Supergravity Group, Department
of Physics, Faculty of Science, Chulalongkorn University, 254 Phayathai Road, Pathumwan, Bangkok 10330, Thailand
\end{center}
E-mail: $^a$parinya.ka@hotmail.com \\
E-mail: $^b$keima.tham@gmail.com \vspace{1 cm}\\
\begin{abstract}
We study matter-coupled $N=3$ gauged supergravity in four dimensions
with various semisimple gauge groups. When coupled to $n$ vector
multiplets, the gauged supergravity contains $3+n$ vector fields and
$3n$ complex scalars parametrized by $SU(3,n)/SU(3)\times
SU(n)\times U(1)$ coset manifold. Semisimple gauge groups take the
form of $G_0\times H\subset SO(3,n)\subset SU(3,n)$ with $H$ being a
compact subgroup of $SO(n+3-\textrm{dim}(G_0))$. The $G_0$ groups
considered in this paper are of the form $SO(3)$, $SO(3,1)$,
$SO(2,2)$, $SL(3,\mathbb{R})$ and $SO(2,1)\times SO(2,2)$. We find
that $SO(3)\times SO(3)$, $SO(3,1)$ and $SL(3,\mathbb{R})$ gauge
groups admit a maximally supersymmetric $AdS_4$ critical point. The
$SO(2,1)\times SO(2,2)$ gauge group admits a supersymmetric
Minkowski vacuum while the remaining gauge groups admit both
half-supersymmetric domain wall vacua and $AdS_4$ vacua with
completely broken supersymmetry. For the $SO(3)\times SO(3)$ gauge
group, there exists another supersymmetric $N=3$ $AdS_4$ critical
point with $SO(3)_{\textrm{diag}}$ symmetry. We explicitly give a
detailed study of various holographic RG flows between $AdS_4$
critical points, flows to non-conformal theories and supersymmetric
domain walls in each gauge group. The results provide gravity duals
of $N=3$ Chern-Simons-Matter theories in three dimensions.
\end{abstract}
\newpage
\section{Introduction}
The AdS/CFT correspondence has attracted a lot of attentions since
its original proposal in \cite{maldacena}. The correspondence
provides a duality relation between a gravity theory in $AdS_{d+1}$
space and a strongly coupled conformal field theory in $d$
dimensions. The correspondence has also been extended to the case of
non-conformal field theories in the form of the domain wall/quantum
field theory (DW/QFT) correspondence \cite{DW/QFT_townsend}. These
provide a useful tool to understand strongly coupled gauge theories
in various space-time dimensions.
\\
\indent AdS$_4$/CFT$_3$ correspondence is particularly interesting
in many aspects. In M-theory, $AdS_4\times X^7$ geometries, with
$X^7$ being an internal compact $7$-dimensional manifold, arise
naturally from a near horizon limit of M2-brane configurations.
AdS$_4$/CFT$_3$ correspondence is then expected to shed some light
on the dynamics of a strongly coupled worldvolume theory on
M2-branes \cite{BL,ABJM}. And, more recently, the correspondence has
also been applied to condensed matter physics systems, see for
example
\cite{Holographi_Superconductor1,Holographi_Superconductor2,Holographi_Superconductor3}.
\\
\indent As in other dimensions, working in lower-dimensional gauged
supergravity has proved to be useful and efficient. In the
lower-dimensional point of view, the $AdS_4\times X^7$ geometries
are identified with the vacua of the scalar potential in the gauged
supergravity theory, and the isometries of the internal manifold correspond
to the gauge symmetry or its unbroken subgroup at the $AdS_4$ vacua. For the case of
$X^7=S^7$, the resulting $AdS_4\times S^7$ geometry preserves
maximal supersymmetry. The effective gauged supergravity in this
case is the maximal $N=8$ $SO(8)$ gauged supergravity in four
dimensions constructed in \cite{N8_4D_deWit}. The holographic study
within this gauged supergravity has been investigated in many
previous works, see for example
\cite{Ahn_4D_flow,Flow_in_N8_4D,4D_G2_flow,Warner_M2_flow,Warner_Fisch}.
These results give a description of the deformations leading to various types of RG
flows in the dual superconformal field theories (SCFTs) in three
dimensions.
\\
\indent For $N>2$ supersymmetry, there is a unique non-maximal $AdS_4$ solution
from a compactification of eleven-dimensional supergravity with unbroken $N=3$ supersymmetry in four
dimensions \cite{N3_compact1}. In this case, the internal manifold
is a tri-sasakian $N^{010}$ with $SU(2)\times SU(3)$ isometry. The corresponding
Kaluza-Klein spectrum has been given in \cite{N3_spectrum1}, and the
structure of $N=3$ multiplets has been further investigated in
\cite{N3_spectrum2}. The properties of the possible dual SCFT to
this background in term of Chern-Simons-Matter theory with $SU(3)$
flavor symmetry has been proposed in
\cite{dual_N3_multiplet,Ring_N3_superfield}. The gravity dual of
this $N=3$ SCFT has been studied in many references, see for example
\cite{AdS_CFT_Tri-Sasakian,D6_AdS4_CP3,gravity_dual_CSM_theory,N3_AdS_CFT1,N3_AdS_CFT2,N3_AdS_CFT3}.
In these results, the four-dimensional scalar potentials, encoding
various deformations of the dual SCFT, have been obtained from
compactifications of eleven-dimensional supergravity restricted to
particular field configurations.
\\
\indent It has been pointed out in \cite{N3_spectrum1} and
\cite{N3_spectrum2} that $AdS_4\times N^{010}$ compactification can
be described by an effective theory in the form of $N=3$,
$SO(3)\times SU(3)$ gauged supergravity coupled to eight vector
multiplets constructed in
\cite{N3_Ferrara,N3_Ferrara2,Castellani_book}. Many supersymmetric
deformations of the maximally supersymmetric $AdS_4$ critical point
including a new $AdS_4$ critical point with $SO(3)\times U(1)$
symmetry have been identified in a recent work \cite{N3_SU2_SU3}.
The eleven-dimensional configurations corresponding to these gravity
solutions might be obtained by a consistent reduction ansatz, to be
explicitly identified.
\\
\indent Apart from a simple compact gauge group studied in \cite{N3_SU2_SU3}, it is natural to consider other types of gauge groups. As in other matter-coupled supergravity, there are many
possible gauged groups for $N=3$ gauged supergravity coupled to $n$
vector multiplets, the only existing matter in $N=3$ supersymmetry.
These gauge groups are in general subgroups of the global, duality, symmetry
group $SU(3,n)$. In this paper, we will consider $N=3$ gauged
supergravity coupled to $n$ vector multiplets with compact and
non-compact gauge groups $\tilde{G}\subset SO(3,n)\subset SU(3,n)$.
In each gauge group, we will study the scalar potential restricted
on scalar submanifolds, which are invariant under particular subgroups of the full gauge group under consideration, and
identify supersymmetric vacua as well as possible RG flow solutions
describing supersymmetric deformations in the dual gauge theories in
three dimensions.
\\
\indent The paper is organized as follow. In section \ref{N3theory},
$N=3$ gauged supergravity coupled to $n$ vector multiplets is
reviewed along with possible semisimple gauge groups allowed by
supersymmetry. The scalar potential of each gauge group is
investigated separately in subsequent sections in which possible
supersymmetric vacua in the form of an $AdS_4$ or a domain wall for
different scalar submanifolds are classified. Conclusions and
comments on the results are presented in section \ref{conclusions}.

\section{$N=3$ gauged supergravity with compact and non-compact gauge groups}\label{N3theory}
We begin with a review of $N=3$ gauged supergravity in four
space-time dimensions constructed in
\cite{N3_Ferrara,N3_Ferrara2,Castellani_book}. We will
closely follow most of the notations in \cite{N3_Ferrara} but in the mostly plus
metric signature $(-+++)$.
\\
\indent $N=3$ supersymmetry in four dimensions contains twelve
supercharges. Apart from the supergravity multiplet, the only matter
multiplets are in the form of vector multiplets. The supergravity
multiplet contains the following field content
\begin{equation}
(e^a_\mu, \psi_{\mu A}, A_{\mu A}, \chi)
\end{equation}
which are given respectively by a graviton $e^a_\mu$, three
gravitini $\psi_{\mu A}$, three vectors $A_{\mu A}$ and one spinor
field $\chi$. Indices $A,B,\ldots=1,2,3$ denote the $SU(3)_R$
R-symmetry triplets while $\mu,\nu,\ldots =0,\ldots,3$ and
$a,b,\ldots=0,\ldots,3$ are respectively space-time and tangent
space indices. Throughout the paper, spinor indices will not be
shown explicitly.
\\
\indent Each of the $n$ vector multiplets has one vector field, four
spinor fields which are a triplet and a singlet of $SU(3)_R$, and
three complex scalars
\begin{equation}
(A_\mu, \lambda_A, \lambda, z_A)^i
\end{equation}
with indices $i,j,\ldots =1,\ldots, n$ labeling each of the vector
multiplets. All spinors are subject to the chirality projection
conditions
\begin{eqnarray}
\psi_{\mu A}&=&\gamma_5\psi_{\mu A},\qquad \chi=\gamma_5\chi,\qquad
\lambda_A=\gamma_5\lambda_A,\qquad
\lambda=-\gamma_5\lambda,\nonumber \\
\psi_\mu^A&=&-\gamma_5\psi_\mu^A, \qquad
\lambda^A=-\gamma_5\lambda^A\, .
\end{eqnarray}
\indent When coupled to $n$ vector multiplets, the supergravity
theory consists of $3n$ complex or $6n$ real scalar fields
$z_A^{\phantom{A}i}$ parametrized by the coset space
$SU(3,n)/SU(3)\times SU(n)\times U(1)$. The scalars can be
parametrized by the coset representative
$L(z)_\Lambda^{\phantom{\Lambda}\Sigma}$ which transforms under the
global $G=SU(3,n)$ and the local $H=SU(3)\times SU(n)\times U(1)$
symmetries by left and right multiplications, respectively.
Indices $\Lambda, \Sigma, \ldots=(A,i)$ take the values $1,\ldots,
n+3$. The indices $i,j,\ldots$ are used to label the fundamental
representation of $SU(n)$. The coset representative can be
accordingly split as follow
$L_\Lambda^{\phantom{\Lambda}\Sigma}=(L_\Lambda^{\phantom{\Lambda}A},L_\Lambda^{\phantom{\Lambda}i})$.
Being an element of $SU(3,n)$, its inverse is related to
$L_\Lambda^{\phantom{\Lambda}\Sigma}$ via the relation
\begin{equation}
(L^{-1})_\Lambda^{\phantom{\Lambda}\Sigma}=J_{\Lambda\Pi}J^{\Sigma\Delta}(L_\Delta^{\phantom{\Delta}\Pi})^*
\end{equation}
where $J_{\Lambda\Sigma}$ is an $SU(3,n)$ invariant tensor given by
\begin{equation}
J_{\Lambda\Sigma}=J^{\Lambda\Sigma}=(\delta_{AB},-\delta_{ij}).
\end{equation}
\indent There are $n+3$ vector fields, three from the gravity
multiplet and $n$ from the vector multiplets, which can be written
collectively by a single notation $A_\Lambda=(A_A,A_i)$. Accompanied
by their magnetic dual, the $n+3$ vector fields transform in the
fundamental representation $\mathbf{n+3}$ of the global symmetry
$SU(3,n)$. The Lagrangian consisting of $n+3$ ``electric''
vectors is invariant only under the $SO(3,n)$ subgroup of the
duality symmetry $SU(3,n)$. It has been argued in
\cite{N3_Ferrara} that possible gauge groups are subgroups of
$SO(3,n)$ which transform the vector fields among themselves. When
restricted to $SO(3,n)$, the fundamental, complex, representation of
$SU(3,n)$ split into two fundamental, real, representations of
$SO(3,n)$
\begin{equation}
(\mathbf{3}+\mathbf{n})_{\mathbb{C}}\rightarrow
(\mathbf{3}+\mathbf{n})_{\mathbb{R}}+(\mathbf{3}+\mathbf{n})_{\mathbb{R}}\,
.
\end{equation}
The $(\mathbf{3}+\mathbf{n})_{\mathbb{R}}$ representation of
$SO(3,n)$ in turn will become the adjoint representation of the
gauge group.
\\
\indent When a particular subgroup $\tilde{G}\subset SO(3,n)\subset
SU(3,n)$ is gauged, the $SO(3,n)$ global symmetry of the Lagrangian
is broken to $\tilde{G}$. The gauge field strengths become
non-abelian defined by
\begin{equation}
F_\Lambda=dA_\Lambda+f_{\Lambda}^{\phantom{\Lambda}\Sigma\Gamma}A_\Sigma\wedge
A_{\Gamma}
\end{equation}
where $f_{\Lambda\Sigma}^{\phantom{\Lambda\Sigma}\Gamma}$ are the
structure constants of the gauge group. The gauge generators
$T_\Lambda$ satisfy the $\tilde{G}$ Lie algebra
\begin{equation}
\left[T_\Lambda,T_\Sigma\right]=f_{\Lambda\Sigma}^{\phantom{\Lambda\Sigma}\Gamma}T_\Gamma\,
.
\end{equation}
It should be noted that $\tilde{G}$ needs not be simple, and each
simple factor can have different coupling constants. Furthermore, in
the presence of gaugings, the Mourer-Cartan one-form on the scalar
manifold gets modified by the gauge fields appearing in the covariant
derivative of $L_\Lambda^{\phantom{\Lambda}\Sigma}$
\begin{equation}
\Omega_{\Lambda}^{\phantom{\Lambda}\Pi}=(L^{-1})_\Lambda^{\phantom{\Lambda}\Sigma}dL_\Sigma^{\phantom{\Sigma}\Pi}
+(L^{-1})_\Lambda^{\phantom{\Lambda}\Sigma}f_{\Sigma}^{\phantom{\Sigma}\Omega\Gamma}A_\Omega
L_\Gamma^{\phantom{\Gamma}\Pi}\, .
\end{equation}
In the following, we will omit all of the gauge fields since we are
only interested in supersymmetric solutions with only the metric and
scalars non-vanishing.
\\
\indent Supersymmetry requires that, for any gauge group consistent
with supersymmetry, the tensor
\begin{equation}
f_{\Lambda\Sigma\Gamma}=f_{\Lambda\Sigma}^{\phantom{\Lambda\Sigma}\Gamma'}J_{\Gamma'\Gamma}
=f_{\left[\Lambda\Sigma\Gamma\right]}
\end{equation}
must be totally antisymmetric. The consistency condition can be
satisfied by taking $J_{\Lambda\Sigma}$ to be the Killing form of
the $(n+3)$-dimensional gauge group $\tilde{G}$. Since
$J_{\Lambda\Sigma}$ has indefinite signs of the eigenvalues, the
gauge groups can be both compact and non-compact types. Furthermore,
since $J_{\Lambda\Sigma}$ has three positive eigenvalues but
arbitrarily large number of negative eigenvalues depending on the
number of vector fields, the gauge group can have at most three
compact or at most three non-compact directions.
\\
\indent Among the possible gauge groups, $SO(3)\times H_n$,
$SO(3,1)\times H_{n-3}$ and $SO(2,2)\times H_{n-3}$ groups, with
$H_n$ being an $n$-dimensional compact group, have been pointed out
in \cite{N3_Ferrara} and \cite{N3_symmetry_breaking}. However, the
consistency condition and the global symmetry $SO(3,n)$ in which the
gauge group can be embedded are very similar to the half-maximal
gauged supergravity in seven dimensions constructed in
\cite{Eric_N2_7D}, and a number of possible gauge groups have been
listed in \cite{Eric_N2_7Dmassive}. We then expect that possible
gauge groups of the $N=3$ gauged supergravity considered here should
follow the same structure.
\\
\indent Due to the restriction on the number of compact or non-compact directions of the gauge group mentioned above, all possible semisimple gauge groups accordingly take the
form of $G_0\times H$ with $H$ being a compact group of dimension
$n+3-\textrm{dim}(G_0)$. It has been pointed out in \cite{Eric_N2_7Dmassive} that $G_0$ is a compact or non-compact group taking one of the following forms
\begin{eqnarray}
& &SO(3),\qquad SO(2,2),\qquad SO(3,1),\nonumber \\
& &SO(2,1),\qquad SO(2,1)\times SO(2,2),\qquad SL(3,\mathbb{R}).
\end{eqnarray}
All of these $G_0$ actually give rise to the gauge groups $G_0\times
H$ with $f_{\Lambda\Sigma\Gamma}=f_{[\Lambda\Sigma\Gamma]}$.
Therefore, they are admissible gauge groups of the $N=3$ gauged
supergravity coupled to vector multiplets.
\\
\indent The bosonic Lagrangian of the $N=3$ gauged supergravity,
with all but the metric and scalars vanishing, can be written as
\begin{equation}
e^{-1}\mc{L}=\frac{1}{4}R-\frac{1}{2}P_\mu^{iA}P^\mu_{Ai}-V\, .
\end{equation}
The vielbein $P_i^{\phantom{i}A}$ of the $SU(3,n)/SU(3)\times
SU(n)\times U(1)$ coset are given by the $(A,i)$-component of the
Mourer-Cartan one-form
$\Omega_i^{\phantom{i}A}=(\Omega_A^{\phantom{A}i})^*$. The scalar
potential is written in terms of the ``boosted structure
constants''
\begin{equation}
C^\Lambda_{\phantom{\Lambda}\Pi\Gamma}=L_{\Lambda'}^{\phantom{\Lambda}\Lambda}
(L^{-1})_{\Pi}^{\phantom{\Lambda}\Pi'}(L^{-1})_{\Gamma}^{\phantom{\Lambda}\Gamma'}
f_{\Pi'\Gamma'}^{\phantom{\Pi'\Gamma'}\Lambda'}\qquad
\textrm{and}\qquad
C_\Lambda^{\phantom{\Lambda}\Pi\Gamma}=J_{\Lambda\Lambda'}J^{\Pi\Pi'}J^{\Gamma\Gamma'}
(C^{\Lambda'}_{\phantom{\Lambda}\Pi'\Gamma'})^*
\end{equation}
by the following relation
\begin{eqnarray}
V&=&-2S_{AC}S^{CM}+\frac{2}{3}\mc{U}_A\mc{U}^A+\frac{1}{6}\mc{N}_{iA}\mc{N}^{iA}
+\frac{1}{6}\mc{M}^{iB}_{\phantom{iB}A}\mc{M}_{iB}^{\phantom{iB}A}\nonumber \\
&=&\frac{1}{8}|C_{iA}^{\phantom{iA}B}|^2+\frac{1}{8}|C_i^{\phantom{A}PQ}|^2-\frac{1}{4}
\left(|C_A^{\phantom{A}PQ}|^2-|C_P|^2\right)
\end{eqnarray}
where $C_P=-C_{PM}^{\phantom{PM}M}$. Various tensors appearing
in the above equation are defined by
\begin{eqnarray}
S_{AB}&=&\frac{1}{4}\left(\epsilon_{BPQ}C_A^{\phantom{A}PQ}+\epsilon_{ABC}C_M^{\phantom{M}MC}\right)\nonumber \\
&=&\frac{1}{8}\left(C_A^{\phantom{A}PQ}\epsilon_{BPQ}+C_B^{\phantom{A}PQ}\epsilon_{APQ}\right),\nonumber \\
\mc{U}^A&=&-\frac{1}{4}C_M^{\phantom{A}MA},\qquad \mc{N}_{iA}=-\frac{1}{2}\epsilon_{APQ}C_i^{\phantom{A}PQ},\nonumber \\
\mc{M}_{iA}^{\phantom{iA}B}&=&\frac{1}{2}(\delta_A^BC_{iM}^{\phantom{iM}M}-2C_{iA}^{\phantom{iA}B}).
\end{eqnarray}
\indent Other important ingredients for finding supersymmetric
solutions are supersymmetry transformations of fermions
\begin{eqnarray}
\delta \psi_{\mu A}&=&D_\mu \epsilon_A+S_{AB}\gamma_\mu \epsilon^B,\\
\delta \chi &=&\mc{U}^A\epsilon_A,\\
\delta \lambda_i&=&-P_{i\mu}^{\phantom{i}A}\gamma^\mu \epsilon_A+\mc{N}_{iA}\epsilon^A,\\
\delta
\lambda_{iA}&=&-P_{i\mu}^{\phantom{i}B}\gamma^\mu\epsilon_{ABC}\epsilon^C
+\mc{M}_{iA}^{\phantom{iA}B}\epsilon_B\, .
\end{eqnarray}
The covariant derivative on the supersymmetry parameter $\epsilon_A$
is defined by
\begin{equation}
D\epsilon_A=d\epsilon_A+\frac{1}{4}\omega^{ab}\gamma_{ab}\epsilon_A+Q_A^{\phantom{A}B}\epsilon_B+\frac{1}{2}nQ
\epsilon_A\,
.
\end{equation}
$Q_A^{\phantom{A}B}$ and $Q$ are the $SU(3)\times U(1)$ composite connections. These
connections and the corresponding ones for $SU(n)$,
$Q_i^{\phantom{i}j}$, can be obtained from $(A,B)$ and $(i,j)$ components of the
Mourer-Cartan one-form
\begin{equation}
\Omega_A^{\phantom{A}B}=Q_A^{\phantom{A}B}-n\delta^B_AQ,\qquad
\Omega_i^{\phantom{i}j}= Q_i^{\phantom{i}j}+3\delta^j_iQ
\end{equation}
with the property that $Q_A^{\phantom{A}A}=Q_i^{\phantom{i}i}=0$.
\\
\indent We are now in a position to study the scalar
potential in each gauge group and classify the corresponding
vacua.

\section{$SO(3)\times SO(3)$ gauge group}\label{SO3_SO3}
We begin with a simple compact gauge group of the form
$SO(3)\times SO(3)$ with $G_0=SO(3)$ and $H_3=SO(3)$. This gauged supergravity can
be obtained from $N=3$ supergravity coupled to three vector
multiplets. The structure constants are given by
\begin{equation}
f_{\Lambda\Sigma}^{\phantom{\Lambda\Sigma}\Gamma}=(g_1\epsilon_{ABC},g_2\epsilon_{i+3,j+3,k+3}),\qquad
i,j=1,2,3\, .
\end{equation}
\\
\indent In this case, there are $18$ scalars parametrized by the
$SU(3,3)/SU(3)\times SU(3)\times U(1)$ coset manifold. To
parametrize this manifold and the other related ones needed in
subsequent sections, we introduce the following $6n$ non-compact
generators for a general $SU(3,n)/SU(3)\times SU(n)\times U(1)$
coset
\begin{eqnarray}
\hat{Y}_{iA}&=&e_{i+3,A}+e_{A,i+3}\qquad \textrm{and}\qquad
\tilde{Y}_{iA}=-ie_{i+3,A}+ie_{A,i+3}
\end{eqnarray}
where $i=1,\ldots, n$ and $(e_{\Lambda\Sigma})_{\Gamma\Delta}=\delta_{\Lambda\Gamma}\delta_{\Sigma\Delta}$.

\subsection{$AdS_4$ vacua and RG flows with $SO(3)$ symmetry}
We first consider scalars which are singlets of
$SO(3)_{\textrm{diag}}\subset SO(3)\times SO(3)$. The $18$ scalars
transform in representations
$(\mathbf{3},\bar{\mathbf{3}})_{-2}+(\bar{\mathbf{3}},\mathbf{3})_2$
of the local $SU(3)\times SU(3)\times U(1)$. From now on, we will
neglect all the $U(1)$ charges for simplicity since they will not play any important role. With the
embedding of $SO(3)$ in $SU(3)$ such that $\mathbf{3}\rightarrow
\mathbf{3}$ and $\bar{\mathbf{3}}\rightarrow \mathbf{3}$, there are
two $SO(3)_{\textrm{diag}}$ singlets among the $18$ scalars
according to the decomposition
\begin{equation}
\mathbf{3}\times \mathbf{3}+\mathbf{3}\times
\mathbf{3}=(\mathbf{1}+\mathbf{3}+\mathbf{5})+(\mathbf{1}+\mathbf{3}+\mathbf{5}).\label{SO3D_decom}
\end{equation}
These singlets correspond to the following $SU(3,3)$ non-compact
generators
\begin{equation}
Y_1=\hat{Y}_{11}+\hat{Y}_{22}+\hat{Y}_{33},\qquad
Y_2=\tilde{Y}_{11}+\tilde{Y}_{22}+\tilde{Y}_{33}\, .
\end{equation}
The coset representative can be parametrized by
\begin{equation}
L=e^{\Phi_1 Y_1}e^{\Phi_2 Y_2}\, .
\end{equation}
The scalar potential is computed to be
\begin{eqnarray}
V&=&-\frac{3}{32}\cosh(2\Phi_2)\left[4\cosh(2\Phi_1)[1+\cosh(2\Phi_1)\cosh(2\Phi_2)]^2g_1^2\right.\nonumber \\
&
&\left.
+2\sinh(2\Phi_1)\left[\cosh(4\Phi_1)-3+2\cosh^2(2\Phi_1)\cosh(4\Phi_2)\right]g_1g_2\right.\nonumber \\
& &
\left.+4\cosh(2\Phi_1)[\cosh(2\Phi_1)
\cosh(2\Phi_2)-1]^2g_2^2\right].\label{SO3_potential}
\end{eqnarray}
We find that this potential admits two supersymmetric $AdS_4$
critical points. The first one occurs at $\Phi_1=\Phi_2=0$ with the
cosmological constant and the $AdS_4$ radius given by
\begin{equation}
V_0=-\frac{3}{2}g_1^2,\qquad L^2=-\frac{3}{2V_0}=\frac{1}{g_1^2}\, .
\end{equation}
Another critical point is given by
\begin{eqnarray}
\Phi_1&=&\frac{1}{2}\ln\left[\frac{g_2-g_1}{g_2+g_1}\right],\qquad
\Phi_2=0,\nonumber \\
V_0&=&-\frac{3g_1^2g_2^2}{2(g_2^2-g_1^2)},\qquad
L^2=\frac{g_2^2-g_1^2}{g_1^2g_2^2}\, .
\end{eqnarray}
It should be noted that reality of $\Phi_1$ requires that
$g_2^2-g_1^2>0$, so the critical point is $AdS_4$ with $V_0<0$.
\\
\indent At the trivial critical point with all scalars vanishing and
$SO(3)\times SO(3)$ symmetry unbroken, all scalars have the same mass
$m^2L^2=-2$ corresponding to the dual operators of dimensions
$\Delta=1,2$ in the dual $N=3$ SCFT. At the $SO(3)_{\textrm{diag}}$
critical point, we can compute the scalar masses as shown in table
\ref{table1}. All masses satisfy the BF bound as expected for a
supersymmetric critical point. Furthermore, there are three massless
Goldstone bosons from the $SO(3)\times SO(3)\rightarrow SO(3)$
symmetry breaking.
\begin{table}[h]
\centering
\begin{tabular}{|c|c|c|}
  \hline
  $SO(3)_{\textrm{diag}}$ representations & $m^2L^2\phantom{\frac{1}{2}}$ & $\Delta$  \\ \hline
  $\mathbf{1}$ & $4$, $-2$ & $4$, $(1,2)$  \\
  $\mathbf{3}$ & $0_{(\times 3)}$, $-2_{(\times 3)}$  & $3$, $(1,2)$ \\
  $\mathbf{5}$ & $-2_{(\times 10)}$  & $(1,2)$ \\
  \hline
\end{tabular}
\caption{Scalar masses at the $N=3$ supersymmetric $AdS_4$ critical
point with $SO(3)_{\textrm{diag}}$ symmetry and the corresponding
dimensions of the dual operators in $SO(3)\times SO(3)$ gauge
group}\label{table1}
\end{table}
\\
\indent To check for the unbroken supersymmetry and set up BPS
equations for studying supersymmetric domain wall solutions, we
consider supersymmetry transformations of $\chi$, $\lambda_i$,
$\lambda_{iA}$ and $\psi_{\mu A}$. The four-dimensional metric is
taken to be
\begin{equation}
ds^2=e^{2A(r)}dx^2_{1,2}+dr^2,
\end{equation}
and the two scalars $\Phi_{1,2}$ only depend on $r$. $\delta\chi=0$
equations are identically satisfied since $C_{M}^{\phantom{M}MA}=0$
in the present case. We will use Majorana representation for gamma
matrices in which all of the gamma matrices $\gamma^a$ are real. The
chirality matrix $\gamma_5=i\gamma^0\gamma^1\gamma^2\gamma^3$ is
then purely imaginary. This implies that $\epsilon_A$ and
$\epsilon^A$ are related by a complex conjugation, $\epsilon_A=(\epsilon^A)^*$.
\\
\indent In the following analysis, we will use the same procedure as
in \cite{N3_SU2_SU3}. With the projection condition
\begin{equation}
\gamma^{\hat{r}}\epsilon_A=e^{i\Lambda}\epsilon^A
\end{equation}
where $e^{i\Lambda}$ is a phase factor, the equations for
$\delta\lambda_i=0$ and $\delta \lambda_{iA}=0$ reduce to two
equations
\begin{eqnarray}
e^{i\Lambda}[\cosh(2\Phi_2)\Phi_1'\pm
i\Phi_2']&=&-\frac{1}{2}\left[\sinh(2\Phi_1)+i\cosh(2\Phi_1\sinh(2\Phi_2))\right]\times
\nonumber \\
& &\left[\cosh\Phi_2(g_1\cosh\Phi_1+g_2\cosh\Phi_1)\right.\nonumber \\
& &\left.-i\sinh
\Phi_2(g_1\sinh\Phi_1+g_2\cosh\Phi_1)\right]\label{dLambda_BPS_SO3}
\end{eqnarray}
where $'\equiv \frac{d}{dr}$.
\\
\indent For this particular coset representative consisting of only $SO(3)_{\textrm{diag}}$ singlets, $S_{AB}$ is
diagonal with
\begin{equation}
S_{AB}=\frac{1}{2}\mc{W}\delta_{AB}
\end{equation}
where the ``superpotential'' $\mc{W}$ is given by
\begin{eqnarray}
\mc{W}&=&-\left[\cosh\Phi_1
\cosh\Phi_2-i\sinh\Phi_1\sinh\Phi_2\right]\left[\cosh\Phi_1\cosh\Phi_2+i\sinh\Phi_1\sinh\Phi_2\right]^2g_1\nonumber
\\
& &+\left[\sinh\Phi_1
\cosh\Phi_2-i\cosh\Phi_1\sinh\Phi_2\right]\left[\sinh\Phi_1\cosh\Phi_2+i\cosh\Phi_1\sinh\Phi_2\right]^2g_2\,
.\nonumber \\
\end{eqnarray}
With this, $\delta \psi_{\mu A}=0$ equations for $\mu=0,1,2$ become
\begin{equation}
A'e^{i\Lambda}+\mc{W}=0\, .\label{delta_psi_SO3}
\end{equation}
By writing $\mc{W}=|\mc{W}|e^{i\omega}$ and separating the real and
imaginary parts of \eqref{delta_psi_SO3}, we find
\begin{eqnarray}
A'+\frac{1}{2}|\mc{W}|(e^{i\omega-i\Lambda}+e^{-i\omega+i\Lambda})&=&0,\\
\frac{1}{2}|\mc{W}|(e^{i\omega-i\Lambda}-e^{-i\omega+i\Lambda})&=&0
\end{eqnarray}
where $W=|\mc{W}|$ will play the role of the ``real
superpotential''. The second equation gives $e^{i\Lambda}=\pm
e^{i\omega}$.
\\
\indent Equation \eqref{dLambda_BPS_SO3} implies $\Phi_2'=0$.
Consistency with the field equations requires that $\Phi_2=0$. We
then set $\Phi_2=0$ in the remaining analysis. Furthermore, setting
$\Phi_2=0$ gives a real $\mc{W}$ since $\omega=0$. In this case, we
simply have $e^{i\Lambda}=\pm 1$, and the BPS equations
\eqref{dLambda_BPS_SO3} and \eqref{delta_psi_SO3} become
\begin{eqnarray}
\Phi_1'&=&\mp\sinh\Phi_1\cosh\Phi_1(g_1\cosh\Phi_1+g_2\sinh\Phi_1),\label{SO3_eq1}\\
A'&=&\pm(g_1\cosh^3\Phi_1+g_2\sinh^3\Phi_1).\label{SO3_eq2}
\end{eqnarray}
These equations admit precisely two $AdS_4$ solutions with $N=3$
supersymmetry identified previously. The corresponding Killing
spinors could be obtained from $\delta \psi_{rA}=0$ which eventually
gives, as in many other cases,
$\epsilon_A=e^{\frac{A}{2}}\epsilon^{(0)}_A$ for constant spinors
$\epsilon^{(0)}_A$ satisfying $\gamma^r\epsilon^{(0)}_A=\pm
\epsilon^{(0)A}$.
\\
\indent It should also be noted that equations \eqref{SO3_eq1} and
\eqref{SO3_eq2} are similar to those studied in \cite{N3_SU2_SU3}
within the $N=3$ gauged supergravity with $SO(3)\times SU(3)$ gauge
group. The solution interpolating between the two supersymmetric
$AdS_4$ critical points can be found similarly. The upper signs will
be chosen in order to identify the UV critical point at $\Phi_1=0$
with $r\rightarrow \infty$. The resulting solution is given by
\begin{eqnarray}
g_1g_2r&=&2g_1\tan^{-1}e^{\Phi_1}+g_2\ln\left[\frac{e^{\Phi_1}+1}{e^{\Phi_1}-1}\right]\nonumber
\\
&
&-2\sqrt{g_2^2-g_1^2}\tanh^{-1}\left[e^{\Phi_1}\sqrt{\frac{g_2+g_1}{g_2-g_1}}\right],\\
A&=&\Phi_1-\ln(1-e^{4\Phi_1})+\ln\left[(e^{2\Phi_1}+1)g_1+(e^{\Phi_1}-1)g_2\right]
\end{eqnarray}
where we have omitted all irrelevant additive integration constants.
\\
\indent As $r\rightarrow \infty$, we find
\begin{equation}
\Phi\sim e^{-g_1r}\sim e^{-\frac{r}{L_{UV}}},\qquad A\sim g_1r\sim
\frac{r}{L_{UV}}\, .
\end{equation}
This implies that the flow is driven by a relevant operator of
dimension $\Delta=1,2$ in the UV. In the IR as $r\rightarrow
-\infty$, we find
\begin{equation}
\Phi_1\sim e^{\frac{g_1g_2r}{\sqrt{g_2^2-g_1^2}}}\sim
e^{\frac{r}{L_{IR}}},\qquad A\sim
\frac{g_1g_2r}{\sqrt{g_2^2-g_1^2}}\sim \frac{r}{L_{IR}}
\end{equation}
which shows that the operator dual to $\Phi_1$ becomes irrelevant
with dimension $\Delta=4$. This precisely agrees with the scalar masses given previously.
\\
\indent Other interesting IR behaviors of the above solution are
flows to large values of $|\Phi_1|$. These correspond to flows from
conformal field theories, identified with the $AdS_4$ critical points, to non-conformal gauge theories in the IR.
As $\Phi_1\rightarrow \infty$, the above solution gives
\begin{eqnarray}
& &\Phi_1\sim -\frac{1}{3}\ln \left[r(g_1+g_2)+C\right],\qquad
A\sim-\Phi_1,\nonumber
\\
& &ds^2=[r(g_1+g_2)+C]^{\frac{2}{3}}dx^2_{1,2}+dr^2\, .
\end{eqnarray}
where $C$ is a constant that can be removed by shifting the
coordinate $r$.
\\
\indent For $\Phi_1\rightarrow -\infty$, we find
\begin{eqnarray}
& &\Phi_1\sim \frac{1}{3}\ln \left[r(g_1-g_2)+C\right],\qquad
A\sim\Phi_1,\nonumber
\\
& &ds^2=[r(g_1-g_2)+C]^{\frac{2}{3}}dx^2_{1,2}+dr^2\, .
\end{eqnarray}
In the above solutions, there is a singularity at $r\sim
-\frac{C}{g_1\pm g_2}$. However, the singularity is physically
acceptable according to the criterion of \cite{Gubser_singularity}
since the potential is bounded above as can be checked from
\eqref{SO3_potential} that
\begin{equation}
V(\Phi_1\rightarrow \pm \infty,\Phi_2=0)\rightarrow -(g_1\pm
g_2)^2\infty\, .
\end{equation}

\subsection{RG flows with $SO(2)\times SO(2)$ symmetry}
We now move to a scalar submanifold invariant under $SO(2)_{\textrm{diag}}\subset SO(2)\times SO(2)\subset SO(3)\times SO(3)$ symmetry. There are six singlets corresponding to $SU(3,3)$ noncompact generators
\begin{eqnarray}
Y_1&=&\hat{Y}_{33},\qquad Y_2=\tilde{Y}_{33},\qquad Y_3=\hat{Y}_{11}+\hat{Y}_{22},\nonumber \\
Y_4&=&\tilde{Y}_{11}+\tilde{Y}_{22},\qquad
Y_5=\hat{Y}_{21}-\hat{Y}_{12},\qquad Y_6=\tilde{Y}_{21}-\tilde{Y}_{12}\, .
\end{eqnarray}
The coset representative can be parametrized by
\begin{equation}
L=e^{\Phi_1Y_1}e^{\Phi_2Y_2}e^{\Phi_3Y_3}e^{\Phi_4Y_4}e^{\Phi_5Y_5}e^{\Phi_6Y_6}\, .\label{L_SO2D}
\end{equation}
The scalar potential turns out to be far more complicated than the $SO(3)$ singlet scalars. We will present the results for some consistent truncations of the full potential.
\\
\indent
We first give the result for $SO(2)\times SO(2)$ singlet scalars. These scalars correspond to $\Phi_1$ and $\Phi_2$. The scalar potential is given by
\begin{equation}
V=-\frac{1}{2}g_1^2e^{-2\Phi_1}\left[e^{2\Phi_1}+(1+e^{4\Phi_1})\cosh(2\Phi_2)\right].\label{V_SO3_SO3_SO2}
\end{equation}
It is clearly seen that this potential admits only a critical point
at $\Phi_1=\Phi_2=0$ which is the $SO(3)\times SO(3)$ critical point.
\\
\indent By using the same projector as in the previous case, we can
set up the relevant BPS equations as follow. In this case, the
matrix $S_{AB}$ is given by
\begin{equation}
S_{AB}=\frac{1}{2}\textrm{diag}(\mc{W}_1,\mc{W}_1,\mc{W}_2)
\end{equation}
where
\begin{eqnarray}
\mc{W}_1&=&-g_1\cosh\Phi_1\cosh\Phi_2, \nonumber \\
\mc{W}_2&=&-g_1(\cosh\Phi_1\cosh\Phi_2+i\sinh\Phi_1\sinh\Phi_2).
\end{eqnarray}
It should be noted that, when $\Phi_1=0$ or $\Phi_2=0$, $\mc{W}_1$
and $\mc{W}_2$ coincide. For $\Phi_1\neq 0$ and $\Phi_2\neq 0$, it
turns out that $\mc{W}_2$ provides the true superpotential in term
of which the scalar potential \eqref{V_SO3_SO3_SO2} can be written
as
\begin{equation}
V=\frac{1}{2}G^{\alpha\beta}\frac{\pd |\mc{W}_2|}{\pd
\Phi_\alpha}\frac{\pd |\mc{W}_2|}{\pd
\Phi_\beta}-\frac{3}{2}|\mc{W}_2|^2\, .
\end{equation}
With the scalar kinetic terms
\begin{equation}
-\frac{1}{2}P^{Ai}_\mu
P^\mu_{iA}=-\frac{1}{2}\left[\cosh^2(2\Phi_2)\Phi_1'^2+\Phi_2'^2\right],
\end{equation}
we find $G_{\alpha\beta}=\textrm{diag}(-\cosh^2(2\Phi_2),-1)$, and
$G^{\alpha\beta}$ is the inverse of $G_{\alpha\beta}$ with
$\Phi_\alpha=(\Phi_1,\Phi_2)$.
\\
\indent The BPS equations coming from $\delta\psi_{\mu A}=0$,
$\mu=0,1,2$, become
\begin{equation}
A'=\mp|\mc{W}_2|=\pm\frac{1}{2}g_1\sqrt{2+2\cosh(2\Phi_1)\cosh(2\Phi_2)}
\end{equation}
and $e^{i\Lambda}=\pm e^{i\omega}$ with
$\mc{W}_2=|\mc{W}_2|e^{i\omega}$. It should also be noted that for
$\Phi_1\neq 0$ and $\Phi_2\neq 0$, only the supersymmetry corresponding
to $\epsilon_3$ can be preserved since we need to set $\epsilon_{1,2}=0$ in the $\delta\psi_{\mu A}$ equations. Therefore, together with the
$\gamma^r$ projection, the solution will preserve only two
supercharges or $N=1$ Poincare supersymmetry in three dimensions.
\\
\indent The conditions $\delta \lambda_{iA}=0$ are identically
satisfied for $\epsilon_{1,2}=0$ while $\delta\lambda_i=0$ equations
give
\begin{eqnarray}
\left[e^{i\Lambda}\left[\cosh(2\Phi_2)\Phi_1'+i\Phi_2'\right]+g_1(\sinh\Phi_1\cosh\Phi_2-i\cosh\Phi_1\sinh\Phi_2)\right]
\epsilon^3=0\,
.
\end{eqnarray}
This will give the flow equations for $\Phi_1$ and $\Phi_2$. Using
the above result for $e^{i\Lambda}=\pm e^{i\omega}$, it can be
verified that the flow equations can be written as
\begin{equation}
\Phi_\alpha'=\pm G^{\alpha\beta}\frac{\pd |\mc{W}_2|}{\pd
\Phi_\beta}
\end{equation}
or explicitly
\begin{eqnarray}
\Phi_1'&=&\mp \frac{\sinh(2\Phi_1)\textrm{sech}(2\Phi_2)g_1}{\sqrt{2+\cosh(2\Phi_1)\cosh(2\Phi_2)}},\nonumber \\
\Phi_2'&=&\mp
\frac{\cosh(2\Phi_1)\sinh(2\Phi_2)g_1}{\sqrt{2+\cosh(2\Phi_1)\cosh(2\Phi_2)}}\,
.
\end{eqnarray}
We are not able to solve the above equations completely, but by
combining the two equations, we find a relation between
$\Phi_1$ and $\Phi_2$
\begin{equation}
\coth(2\Phi_2)=\frac{e^{2\Phi_1}}{2-2e^{4\Phi_1}}\, .
\end{equation}
The full flow solution would require some numerical analysis. In the
following, we will simply give the asymptotic behaviors at $\Phi_{1,2}\sim 0$ and large
$|\Phi_\alpha|$.
\\
\indent Identifying $r\rightarrow \infty$ as the UV fixed point, we find
\begin{equation}
\Phi_1\sim\Phi_2\sim e^{-g_1r}
\end{equation}
As $\Phi_2\rightarrow \pm\infty$, we find
\begin{eqnarray}
& &\Phi_1\sim \Phi_0,\qquad \Phi_2\sim \mp\ln(g_1r)\nonumber \\
& &ds^2=r^2dx^2_{1,2}+dr^2
\end{eqnarray}
where $\Phi_0$ is a constant. For convenience, we have put the
singularity at $r=0$ by choosing an integration constant. \\
\indent For $\Phi_1\rightarrow \pm\infty$, the solution becomes
\begin{eqnarray}
& &\Phi_1\sim\mp \ln(g_1r),\qquad \Phi_2\sim \Phi_0,\nonumber \\
& &ds^2=r^2dx^2_{1,2}+dr^2\, .
\end{eqnarray}
All of these flows give $V\rightarrow-\infty$ and are physical.
\\
\indent As noted before for $\Phi_1$ or $\Phi_2$ vanishing, the
eigenvalues of $S_{AB}$ degenerate $\mc{W}_1=\mc{W}_2$, and the BPS
equations coming from $\delta\lambda_i=0$ and $\delta\lambda_{iA}=0$
are identical. The resulting equations for $\Phi_1=0$ and $\Phi_2=0$
cases turn out to be symmetric. In the following, we will set $\Phi_2=0$ for
definiteness. The flow equations reduce to
\begin{eqnarray}
\Phi_1'&=&- g_1\sinh\Phi_1,\nonumber \\
A'&=&g_1\cosh\Phi_1
\end{eqnarray}
with a simple solutions
\begin{eqnarray}
\Phi_1&=&\pm
\ln\left[\frac{e^{g_1r-C}+1}{e^{g_1r-C}-1}\right],\nonumber \\
A&=&-g_1r+\ln(e^{2g_1r-2C}-1).
\end{eqnarray}
At large $r$, we find $\Phi_1\sim e^{-g_1r}$ and $A\sim g_1r$ which
is the UV $AdS_4$. For $g_1r\sim C$, the solution becomes
\begin{eqnarray}
& &\Phi_1\sim \pm \ln(g_1r-C),\qquad A\sim \ln(g_1r-C),\nonumber \\
& &ds^2=(g_1r-C)^2dx^2_{1,2}+dr^2\, .
\end{eqnarray}
This solution is also physical and preserves $N=3$ Poincare
supersymmetry in three dimensions. We then find two classes of
deformations that break conformal symmetry. One of them with
$\Phi_1$ and $\Phi_2$ non-vanishing breaks $N=3$ supersymmetry to
$N=1$ while the other with $\Phi_1$ or $\Phi_2$ vanishing
preserves $N=3$ supersymmetry. On the other hand, both of them preserve $SO(2)\times SO(2)$ symmetry.

\subsection{RG flows with $SO(2)$ symmetry}
The scalar potential and BPS equations for $SO(2)_{\textrm{diag}}$
singlet scalars are far more complicated than the $SO(2)\times
SO(2)$ case. We will only give the result for a truncation with
$\Phi_2=\Phi_4=\Phi_6=0$. We have verified that this is a consistent
truncation both for the BPS equations and the corresponding field
equations.
\\
\indent In this truncation, $S_{AB}$ is diagonal
\begin{equation}
S_{AB}=\frac{1}{2}\mc{W}\delta_{AB}
\end{equation}
where $\mc{W}$ is real and given by
\begin{eqnarray}
W=\mc{W}&=&-\frac{1}{2}g_1\cosh\Phi_1[1+\cosh(2\Phi_3)]\cosh(2\Phi_5)\nonumber
\\
& &+g_2[1-\cosh(2\Phi_3)\cosh(2\Phi_5)]\sinh\Phi_1\, .
\end{eqnarray}
With the scalar kinetic terms
\begin{equation}
-\frac{1}{2}P_\mu^{iA}P_{Ai}^\mu=-\frac{1}{2}\Phi_1'^2-\frac{1}{4}e^{-4\Phi_5}(1+e^{4\Phi_5})^2\Phi_3'^2
-\Phi_5'^2,
\end{equation}
the scalar potential can be written as
\begin{eqnarray}
V&=&-\frac{1}{2}\frac{\pd W}{\pd \Phi_1}\frac{\pd W}{\pd
\Phi_1}-\frac{e^{4\Phi_5}}{(1+e^{4\Phi_5})^2}\frac{\pd W}{\pd
\Phi_3}\frac{\pd W}{\pd \Phi_3}-\frac{1}{4}\frac{\pd W}{\pd
\Phi_5}\frac{\pd W}{\pd \Phi_5}-\frac{3}{2}W^2\nonumber \\
&=&\frac{1}{32}\left[-4[1+\cosh(2\Phi_3)\cosh(2\Phi_5)]\left[2\cosh(2\Phi_3)\cosh(2\Phi_5)\right.\right.\nonumber
\\
&
&\left.+\cosh(2\Phi_1)[1+3\cosh(2\Phi_3)]\cosh(2\Phi_5)\right]g_1^2\nonumber
\\
&
&-6\left[\cosh(4\Phi_3)+2\cosh^2(2\Phi_3)\cosh(4\Phi_5)-3\right]\sinh(2\Phi_1)g_1g_2,\nonumber
\\
&
&+2[2\cosh(2\Phi_3)\cosh(2\Phi_5)-2]\left[2\cosh(2\Phi_3)\cosh(2\Phi_5)\right.\nonumber
\\
&
&\left.\left.+2\cosh(2\Phi_1)[1-3\cosh(2\Phi_3)\cosh(2\Phi_5)]\right]g_2^2\right].
\end{eqnarray}
All of the BPS equations coming from $\delta\lambda_i=0$ and
$\delta\lambda_{iA}=0$ are solved by the following flow equations
\begin{eqnarray}
\Phi_1'&=&\pm\frac{\pd W}{\pd \Phi_1}\nonumber
\\
&=&\mp\frac{1}{2}\left[g_1[1+\cosh(2\Phi_3)\cosh(2\Phi_5)]\sinh\Phi_1\right.\nonumber \\
& &\left.+g_2\cosh\Phi_1
[\cosh(2\Phi_3)\cosh(2\Phi_5)-1]\right],\\
\Phi_3'&=&\pm\frac{2e^{4\Phi_5}}{(1+e^{4\Phi_5})^2}\frac{\pd W}{\pd
\Phi_3}\nonumber \\
&=&\mp
\frac{e^{2\Phi_5}}{1+e^{4\Phi_5}}\sinh(2\Phi_3)[g_1\cosh\Phi_1+g_2\sinh\Phi_1],\\
\Phi_5'&=&\pm\frac{1}{2}\frac{\pd W}{\pd \Phi_5}\nonumber \\
&=&\mp
\frac{1}{2}\cosh(2\Phi_3)\sinh(2\Phi_5)[g_1\cosh\Phi_1+g_2\sinh\Phi_1],\\
A'&=&\mp W
\end{eqnarray}
after using the projector $\gamma^r\epsilon_A=\pm \epsilon^A$. The
solution to these equations then preserves $N=3$ supersymmetry in
three dimensions. Apart from the trivial critical point with all
$\Phi_i=0$ and the $SO(3)_{\textrm{diag}}$ with $\Phi_5=0$ and
$\Phi_1=\pm\Phi_3=\frac{1}{2}\ln\left[\frac{g_2+g_1}{g_2-g_1}\right]$,
the above equations admit no new critical points.
\\
\indent We are not able to solve the above equations analytically
for general values of $g_1$ and $g_2$. However, for $g_2=g_1$ and
$\Phi_5=0$, an analytic solution can be found
\begin{eqnarray}
A&=&\Phi_1-\frac{1}{2}\ln(e^{4\Phi_1}-1),\nonumber \\
\Phi_3&=&\cosh^{-1}\left[e^{\frac{\Phi_1}{2}}\sqrt{\cosh\Phi_1}\right],\nonumber
\\
g_1r&=&\tan^{-1}
e^{\Phi_1}+\frac{1}{2}\ln\left[\frac{e^{\Phi_1}+1}{e^{\Phi_1}-1}\right].
\end{eqnarray}
This solution describes an RG flow from the trivial $AdS_4$ critical
point to an $N=3$ non-conformal gauge theory in the IR. At
$\Phi_1\sim \Phi_3\sim 0$, the above solution
approaches the UV $AdS_4$
\begin{equation}
\Phi_1\sim e^{-2g_1r},\qquad \Phi_3\sim e^{-g_1r},\qquad A\sim
g_1r\, .
\end{equation}
Near the IR singularity $r\sim 0$, the solution behaves as
\begin{eqnarray}
& &\Phi_1\sim -\ln(g_1r),\qquad \Phi_3\sim\Phi_1,\qquad A\sim
-\Phi_1\sim \ln(g_1r),\nonumber \\
& &ds^2=(g_1r)^2dx^2_{1,2}+dr^2
\end{eqnarray}
for $\Phi_1>0$ and
\begin{eqnarray}
& &\Phi_1\sim \ln(g_1r),\qquad \Phi_3\sim\textrm{constant},\qquad
A\sim
\Phi_1\sim \ln(g_1r),\nonumber \\
& &ds^2=(g_1r)^2dx^2_{1,2}+dr^2
\end{eqnarray}
for $\Phi_1<0$. Both of these singularities give $V\sim -\infty$ and
hence are physical. Therefore, the solution gives a gravity dual of
an RG flow from $N=3$ SCFT with $SO(3)\times SO(3)$ symmetry to
$N=3$ gauge theory with $SO(2)$ symmetry in three dimensions.

\section{$SO(3,1)$ gauge group}\label{SO3_1}
We still work with the $n=3$ case but with $SO(3,1)$ gauge group.
The structure constants in this case are given by $f_{\Lambda\Sigma}^{\phantom{\Lambda\Sigma}\Gamma}=f_{\Lambda\Sigma\Gamma'}J^{\Gamma'\Gamma}$ where
\begin{equation}
f_{\Lambda\Sigma\Gamma}=g(\epsilon_{ABC},\epsilon_{i+3,j+3,A}),
\end{equation}
and $\epsilon_{i+3,j+3,A}$ are totally antisymmetric with
$\epsilon_{345}=\epsilon_{156}=\epsilon_{264}=1$.

\subsection{RG flows with $SO(3)$ symmetry}
We now proceed as in the previous section by considering
$SO(3)\subset SO(3,1)$ singlet scalars. Under this $SO(3)$, the
decomposition of the representation for all $18$ scalars is similar to
\eqref{SO3D_decom} since the $SO(3)$ maximal compact subgroup of
$SO(3,1)$ is embedded in $SO(3,1)$ as a diagonal subgroup of
$SO(3)\times SO(3)\subset SO(3,3)$. Accordingly, there are two
singlets given by the $SU(3,3)$ non-compact generators
\begin{equation}
Y_1=\hat{Y}_{11}-\hat{Y}_{22}+\hat{Y}_{3},\qquad
Y_2=\tilde{Y}_{11}-\tilde{Y}_{22}+\tilde{Y}_{33}\, .
\end{equation}
We then parametrize the coset representative by
\begin{equation}
L=e^{\Phi_1Y_1}e^{\Phi_2Y_2}
\end{equation}
which gives the potential
\begin{eqnarray}
V&=&-\frac{3}{64}g^2e^{-6\Phi_1}\left[2e^{6\Phi_1}\left[13\cosh(2\Phi_1)+3\cosh(6\Phi_1)\right]\cosh(2\Phi_2)
\right. \nonumber \\
&
&\left.+(e^{4\Phi_1}-1)^2\left[(1+e^{4\Phi_1})\cosh(6\Phi_2)-16e^{2\Phi_1}\cosh^2(2\Phi_2)\right]\right].
\end{eqnarray}
This potential admits a trivial critical point at $\Phi_1=\Phi_2=0$
at which the $SO(3,1)$ gauge symmetry is broken to its maximal
compact subgroup $SO(3)$. The values of the cosmological constant and
$AdS_4$ radius are given by
\begin{equation}
V_0=-\frac{3}{2}g^2,\qquad L^2=\frac{1}{g^2}\, .
\end{equation}
Scalar masses are given in table \ref{table2}. We again see that
there are three Goldstone bosons.
\begin{table}[h]
\centering
\begin{tabular}{|c|c|c|}
  \hline
  $SO(3)$ representations & $m^2L^2\phantom{\frac{1}{2}}$ & $\Delta$  \\ \hline
  $\mathbf{1}$ & $4$, $-2$ & $4$, $(1,2)$  \\
  $\mathbf{3}$ & $0_{(\times 3)}$, $-2_{(\times 3)}$  & $3$, $(1,2)$ \\
  $\mathbf{5}$ & $-2_{(\times 10)}$  & $(1,2)$ \\
  \hline
\end{tabular}
\caption{Scalar masses at the $N=3$ supersymmetric $AdS_4$ critical
point with $SO(3)$ symmetry and the corresponding dimensions of the
dual operators in $SO(3,1)$ gauge group}\label{table2}
\end{table}
\\
\indent We also find a non-supersymmetric critical point given by
\begin{eqnarray}
\Phi_1&=&\frac{1}{2}\ln\left[\frac{4\pm\sqrt{7}}{3}\right],\qquad \Phi_2=0,\nonumber \\
V_0&=&-\frac{11}{9}g^2,\qquad L^2=\frac{27}{22g^2}\, .
\end{eqnarray}
This critical point is however unstable since some of the scalar
masses violate the BF bound. All scalar masses are given in table
\ref{table3}.
\begin{table}[h]
\centering
\begin{tabular}{|c|c|}
  \hline
  $SO(3)$ representations & $m^2L^2\phantom{\frac{1}{2}}$   \\ \hline
  $\mathbf{1}$ & $-\frac{168}{11}$, $-\frac{36}{11}$   \\
  $\mathbf{3}$ & $0_{(\times 3)}$, $\left.-\frac{36}{11}\right|_{(\times 3)}$ \\
  $\mathbf{5}$ & $\left.-\frac{24}{11}\right|_{(\times 5)}$, $\left.-\frac{36}{11}\right|_{(\times 5)}$  \\
  \hline
\end{tabular}
\caption{Scalar masses at the non-supersymmetric $AdS_4$ critical
point with $SO(3)$ symmetry in $SO(3,1)$ gauge group}\label{table3}
\end{table}
\\
\indent We now consider possible supersymmetric RG flow solutions
within the $N=3$ $SO(3,1)$ gauged supergravity. Since we have not
found any non-trivial supersymmetric $AdS_4$ critical points in this
gauge group, we will consider only supersymmetric RG flows to
non-conformal theories. Similar to the $SO(3)\times SO(3)$ gauge
group, we find that the BPS equations coming from
$\delta\lambda_i=0$ and $\delta\lambda_{iA}=0$ give rise to the
following equations
\begin{eqnarray}
e^{i\Lambda}\left[\cosh(2\Phi_2)\Phi_1'\pm
i\Phi_2'\right]&=&g\sinh^3\Phi_1\cosh\Phi_2+
\frac{1}{2}g\cosh\Phi_1\left[\sinh(2\Phi_1)\cosh(3\Phi_2)\right.\nonumber
\\
& &\left.
-2i\left[1-2\sinh^2\Phi_1\cosh(2\Phi_2)\right]\sinh\Phi_2\right]
\end{eqnarray}
which again imply $\Phi_2'=0$. Consistency with the second order
field equations requires that $\Phi_2=0$. This gives rise to real
superpotential.
\\
\indent Follow the same procedure as in the previous section with an
appropriate sign choice, we find the relevant BPS equations
\begin{eqnarray}
\Phi_1'&=&\frac{1}{4}e^{-3\Phi_1}g\left(e^{2\Phi_1}+e^{6\Phi_1}-e^{4\Phi_1}-1\right),\\
A'&=&-\frac{1}{4}e^{-3\Phi_1}g\left(1+e^{6\Phi_1}-3e^{2\Phi_1}-3e^{4\Phi_1}\right).
\end{eqnarray}
Since the operator dual to $\Phi_1$ has dimension $\Delta=4$
corresponding to an irrelevant operator, we then expect the $AdS_4$
to appear in the IR of the RG flow driven by $\Phi_1$. The solution to the above equations can be readily found
\begin{eqnarray}
gr&=&\ln\left[\frac{e^{\Phi_1}-1}{e^{\Phi_1}+1}\right]
+\frac{1}{\sqrt{2}}\ln\left[\frac{1+\sqrt{2}e^{\Phi_1}+e^{2\Phi_1}}{\sqrt{2}e^{\Phi_1}-1-e^{2\Phi_1}}\right],\\
A&=&\Phi_1+\ln(e^{2\Phi_1}-1)-\ln(1+e^{4\Phi_1}).
\end{eqnarray}
As $\Phi_1\sim 0$, the solution gives
\begin{equation}
\Phi_1\sim e^{gr}\sim e^{\frac{r}{L}},\qquad A\sim gr\sim
\frac{r}{L}
\end{equation}
which is the $AdS_4$ critical point.
\\
\indent At large $|\Phi_1|$, we find that for $\Phi_1>0$ the
solution behaves as
\begin{eqnarray}
& &\Phi_1\sim-\frac{1}{3}\ln(gr+C),\qquad A\sim -\Phi_1,\nonumber \\
& &ds^2=(gr+C)^{\frac{2}{3}}dx^2_{1,2}+dr^2
\end{eqnarray}
while for $\Phi_1<0$, the solution becomes
\begin{eqnarray}
& &\Phi_1\sim\frac{1}{3}\ln(C-gr),\qquad A\sim \Phi_1,\nonumber \\
& &ds^2=(C-gr)^{\frac{2}{3}}dx^2_{1,2}+dr^2\, .
\end{eqnarray}
Both of these singularities are physical since
\begin{equation}
V(\Phi_1\rightarrow\pm \infty,\Phi_2=0)\rightarrow -\infty\, .
\end{equation}

\subsection{RG flows with $SO(2)$ symmetry}
For $SO(2)$ singlet scalars, the coset representative can be
parametrized by
\begin{equation}
L=e^{\Phi_1Y_1}e^{\Phi_2Y_2}e^{\Phi_3Y_3}e^{\Phi_4Y_4}e^{\Phi_5Y_5}e^{\Phi_6Y_6}
\end{equation}
where the $SU(3,3)$ non-compact generators are defined by
\begin{eqnarray}
Y_1&=&\hat{Y}_{33},\qquad Y_2=\tilde{Y}_{33},\qquad Y_3=\hat{Y}_{11}-\hat{Y}_{22},\nonumber \\ Y_4&=&\tilde{Y}_{11}-\tilde{Y}_{22},\qquad
Y_5=\hat{Y}_{12}+\hat{Y}_{21},\qquad
Y_6=\tilde{Y}_{12}+\tilde{Y}_{21}\, .
\end{eqnarray}
The resulting scalar potential is very complicated. After
making a truncation by setting $\Phi_2=\Phi_4=\Phi_6=0$, we find a
much simpler potential
\begin{eqnarray}
V&=&\frac{1}{8}g^2\left[16\cosh(2\Phi_5)\sinh(2\Phi_1)\sinh(2\Phi_3)-3\cosh(2\Phi_1)[3+\cosh(4\Phi_3)]\right.
\nonumber \\
&
&\left.+2[2+(2-3\cosh(2\Phi_1))\cosh(4\Phi_5)]\sinh^2(2\Phi_3)\right].
\end{eqnarray}
Apart from the trivial critical point, there are no other
supersymmetric critical points from this potential.
\\
\indent We now move to the BPS equations. The $S_{AB}$ matrix in
this truncation is diagonal and proportional to the identity matrix with the superpotential
\begin{equation}
W=-g\cosh\Phi_1+g\cosh(2\Phi_5)\sinh\Phi_1\sin(2\Phi_3).
\end{equation}
As usual, the scalar potential can be written in term of $W$ as
\begin{equation}
V=-\frac{1}{2}\left(\frac{\pd W}{\pd
\Phi_1}\right)^2-e^{4\Phi_5}(1+e^{4\Phi_5})^{-2}\left(\frac{\pd
W}{\pd \Phi_3}\right)^2-\frac{1}{4}\left(\frac{\pd W}{\pd
\Phi_5}\right)^2-\frac{3}{2}W^2\, .
\end{equation}
The flow equations are then given by
\begin{eqnarray}
\Phi_1'&=&\pm \frac{\pd W}{\pd \Phi_1}=\pm\left[
-g\sinh\Phi_1+g\cosh\Phi_1\cosh(2\Phi_5)\sinh(2\Phi_3)\right],\\
\Phi_3'&=&\pm 2e^{4\Phi_5}(1+e^{4\Phi_5})^{-2}\frac{\pd W}{\pd
\Phi_3}\nonumber \\
&=&\pm
\frac{4e^{4\Phi_5}}{(1+e^{4\Phi_5})^2}g\cosh(2\Phi_3)\cosh(2\Phi_5)\sinh\Phi_1,\\
\Phi_5'&=&\pm\frac{1}{2}\frac{\pd W}{\pd \Phi_5}=\pm
g\sinh\Phi_1\sinh(2\Phi_3)\sinh(2\Phi_5),\\
A&=&\mp W\, .
\end{eqnarray}
We are not able to solve these equations analytically. We will
therefore only discuss the asymptotic behaviors of the flow solution
and leave the full solution for a numerical analysis. Near the
$AdS_4$ critical point, we find
\begin{equation}
\Phi_1\sim\Phi_3\sim e^{gr}\sim e^{\frac{r}{L}},\qquad \Phi_5\sim
\textrm{constant},\qquad A\sim gr\sim\frac{r}{L}\, .
\end{equation}
We see that $\Phi_1$ and $\Phi_3$ are dual to irrelevant operators
of dimension $4$ while $\Phi_5$ is dual to a marginal operator.
Actually, $\Phi_5$ is one of the Goldstone bosons.
\\
\indent Near the singularity at large $|\Phi_3|$, we find
$\Phi_5'=0$. In what follows, we will choose $\Phi_5=0$ for simplicity. The
asymptotic behaviors of the flow solution are given by
\begin{eqnarray}
& &\Phi_1\sim \pm\Phi_3\sim \pm \frac{1}{3}\ln\left|C\pm \frac{3}{4}gr\right|,\qquad A\sim \frac{1}{3}\ln\left|C\pm \frac{3}{4}gr\right|,\nonumber \\
& & ds^2=\left(C\pm \frac{3}{4}gr\right)^{\frac{2}{3}}dx^2_{1,2}+dr^2\, .
\end{eqnarray}
It can also be checked that both of these singularities are physical.

\section{$SO(2,2)$ gauge group}\label{SO2_2}
For $n=3$ vector multiplets, there is another possible gauge group
namely $SO(2,2)\sim SO(2,1)\times SO(2,1)$. The structure
constants are given by
\begin{equation}
f_{\Lambda\Sigma}^{\phantom{\Lambda\Sigma}\Gamma}=(g_1\epsilon_{\bar{A}\bar{B}\bar{D}}\eta^{\bar{D}\bar{C}},
g_2\epsilon_{\bar{i}\bar{j}\bar{l}}\eta^{\bar{l}\bar{k}}),\qquad
\bar{A},\bar{B},\ldots =1,2,6,\qquad \bar{i},\bar{j},\ldots=3,4,5
\end{equation}
with $\eta^{\bar{A}\bar{B}}=\textrm{diag}(1,1,-1)$ and
$\eta^{\bar{i}\bar{j}}=\textrm{diag}(1,-1,-1)$.
\\
\indent We will consider the scalar potential for
$SO(2)_\textrm{diag}$ invariant scalars. There are six singlets
parametrized by the coset representative
\begin{equation}
L=e^{\Phi_1(\hat{Y}_{11}+\hat{Y}_{22})}e^{\Phi_2(\tilde{Y}_{11}+\tilde{Y}_{22})}
e^{\Phi_3\hat{Y}_{33}}e^{\Phi_4\tilde{Y}_{33}}
e^{\Phi_5(\hat{Y}_{21}-\hat{Y}_{12})}e^{\Phi_6(\tilde{Y}_{21}-\tilde{Y}_{12})}\,
.
\end{equation}
The scalar potential turns out be much involved. We will only give
the potential for a truncation $\Phi_2=\Phi_4=\Phi_6=0$ for brevity
\begin{eqnarray}
V&=&\frac{1}{16}\left[4 \cosh (2\Phi_{1}) \cosh (2\Phi_{5})
\left[\cosh (2\Phi_{1}) \cosh (2\Phi_{5}) (g_{1}^2-g_{2}^2) +g_{1}^2+g_{2}^2\right]-\right.\nonumber\\
& &2\cosh (2\Phi_{3})\left[g_{1}^2+g_{2}^2+\cosh (2\Phi_{1}) \cosh
(2\Phi_{5}) \right.\nonumber \\
& &\left.\times\left[3 \cosh (2\Phi_{1}) \cosh (2\Phi_{5})
\left(g_{1}^2+g_{2}^2\right)+4 (g_{1}^2-g_{2}^2) \right]\right]\nonumber\\
& & \left.+3 g_{1} g_{2} \sinh (2\Phi_{3}) \left[2 \cosh (4 \Phi_{5})
\cosh^2(2\Phi_{1})+\cosh (4 \Phi_{1})-3\right]\right]
\end{eqnarray}
This potential admits an $AdS_4$ critical point at $\Phi_i=0$,
$i=1,\ldots, 6$ with $V_0=-\frac{1}{2}g_1^2$ and $L^2=\frac{3}{g_1^2}$. This critical point is
however non-supersymmetric. This can be seen by considering the
supersymmetry transformations
\begin{equation}
\delta\lambda_{i}=\delta_{i3}g_1\epsilon^3\qquad \textrm{and}\qquad
\delta\lambda_{iA}=\delta_{i3}g_1(\delta_{A2}\epsilon^1-\delta_{A1}\epsilon^2).
\end{equation}
We see that the only way these variations will vanish is to set
$\epsilon^A=0$, so this critical point breaks all supersymmetries. This critical point is also unstable as can be seen from the scalar masses in table \ref{table5}
\begin{table}[h]
\centering
\begin{tabular}{|c|c|}
  \hline
  $SO(2)\times SO(2)$ representations & $m^2L^2\phantom{\frac{1}{2}}$   \\ \hline
  $(\mathbf{1},\mathbf{1})$ & $-6$, $-6$   \\
  $(\mathbf{2},\mathbf{1})$ & $0_{(\times 2)}$, $\left.-\frac{15}{2}\right|_{(\times 2)}$ \\
  $(\mathbf{1},\mathbf{2})$ & $0_{(\times 2)}$, $\left.-\frac{3g_2^2}{2g_1^2}\right|_{(\times 2)}$  \\
  $(\mathbf{2},\mathbf{2})$ & $\left.-\frac{3}{2}\frac{g_1^2+g_2^2}{g_1^2}\right|_{(\times 8)}$  \\
  \hline
\end{tabular}
\caption{Scalar masses at the non-supersymmetric $AdS_4$ critical
point with $SO(2)\times SO(2)$ symmetry in $SO(2,2)$ gauge group}\label{table5}
\end{table}
\\
\indent On the other hand, a half-supersymmetric vacuum in the form
of a domain wall is possible. Use the domain wall ansatz for the
metric and proceed as in the previous cases, we find a set of very
complicated BPS equations for $SO(2)_{\textrm{diag}}$ singlet scalars. To give an example of this solution, we will consider a simpler case of $SO(2)\times SO(2)$ symmetry. Setting all scalars but $\Phi_3$ and $\Phi_4$ to zero results in a simple scalar potential
\begin{equation}
V=-\frac{1}{2}g_1^2e^{-2\Phi_3}\left[(1+e^{4\Phi_3})\cosh(2\Phi_4)-e^{2\Phi_3}\right].
\end{equation}
The gravitini variations give
\begin{equation}
S_{AB}=\frac{1}{2}\textrm{diag}(\mc{W}_1,\mc{W}_1,\mc{W}_2)
\end{equation}
where
\begin{eqnarray}
\mc{W}_1&=&g_1\sin\Phi_3\cosh\Phi_4,\\
\mc{W}_2&=&g_1\cosh\Phi_4\sinh\Phi_3+ig_1\cosh\Phi_3\sinh\Phi_4\, .
\end{eqnarray}
As in the $SO(3)\times SO(3)$ case, only supersymmetry generated by $\epsilon_3$ is preserved. Carrying out a similar analysis gives the following BPS equations
\begin{eqnarray}
\Phi_3'&=&\pm\cosh^{-2}(2\Phi_4)\frac{\pd W}{\pd\Phi_3}=\pm \frac{g_1\textrm{sech}(2\Phi_4)\sinh(2\Phi_3)}
{\sqrt{2}\sqrt{\cosh(2\Phi_3)\cosh(2\Phi_4)-1}},\\
\Phi_4'&=&\mp\frac{\pd W}{\pd \Phi_4}=\mp\frac{g_1\cosh(2\Phi_3)\sinh(2\Phi_4)}
{\sqrt{2}\sqrt{\cosh(2\Phi_3)\cosh(2\Phi_4)-1}},\\
A'&=&\mp W
\end{eqnarray}
where
\begin{equation}
W=|\mc{W}_2|=\sqrt{2}g_1\sqrt{\cosh(2\Phi_3)\cosh(2\Phi_4)-1}\, .
\end{equation}
From these equations, we immediately see that there is no
supersymmetric $AdS_4$ critical point. We can also solve for $A$ and
$\Phi_3$ as a function of $\Phi_4$ as follow
\begin{eqnarray}
\Phi_3&=&\frac{1}{2}\ln\left[\frac{1}{4}\left[\textrm{csch}(2\Phi_4)\sqrt{10\cosh(4\Phi_4)-6}-2\coth(2\Phi_4)\right]\right],\\
A&=&-\frac{1}{2}\ln \sinh(2\Phi_4)-iF(2i\Phi_4,5)
\end{eqnarray}
where $F$ is the elliptic function of the first kind defined by
\begin{equation}
iF(i\Phi_3,5)=\int_0^{\Phi_3}\frac{d\chi}{\sqrt{1-25\sinh^3\chi}}\, .
\end{equation}
However, we are not able to solve for $\Phi_4(r)$ in a closed form.
\\
\indent For $\Phi_4=0$, $|\mc{W}_1|=|\mc{W}_2|$, we find much simpler BPS equations
\begin{eqnarray}
\Phi_3'&=&\pm g_1\cosh\Phi_3,\\
A'&=&\pm g_1\sinh\Phi_3\, .
\end{eqnarray}
It should be noted that in this case the supersymmetry is enhanced
to $N=3$ as in the $SO(3)\times SO(3)$ case. An analytic solution to
these equations can be completely obtained
\begin{eqnarray}
\Phi_3&=&\ln\tan\left[\frac{g_1r+C}{2}\right],\qquad A=-\ln\sin(g_1r+C),\\
ds^2&=&\sin^{-2}(g_1r+C)dx^2_{1,2}+dr^2\, .\label{SO2_2_DW}
\end{eqnarray}
The solution preserves $N=3$ Poincare supersymmetry in three
dimensions due to the projection $\gamma^r\epsilon_A=\pm\epsilon^A$.
According to the DW/QFT correspondence, this solution should be dual
to a three-dimensional $N=3$ gauge theory.
\\
\indent We end this section by giving a remark on
$SO(2,1)$ gauge group. This gauge group can be obtained by coupling
one vector multiplet to the $N=3$ supergravity and gauging the
theory by using the structure constant
\begin{equation}
f_{\Lambda\Sigma}^{\phantom{\Lambda\Sigma}\Gamma}=g\epsilon_{\bar{A}\bar{B}\bar{D}}\eta^{\bar{D}\bar{C}},\qquad
\bar{A},\bar{B},\ldots =1,2,4,\qquad
\eta_{\bar{A}\bar{B}}=\textrm{diag}(1,1,-1)\, .
\end{equation}
The resulting potential and BPS equations for $SO(2)\subset SO(2,1)$
invariant scalars are the same as the above results for $SO(2,2)$
gauge group with $g_2=0$. Therefore, $SO(2,1)$ gauge group also
admits a non-supersymmetric $AdS_4$ critical point with all scalars
vanishing and a half-supersymmetric domain wall. In particular, the
domain wall with $SO(2)$ symmetry has the same form as the solution
given in \eqref{SO2_2_DW}.

\section{$SL(3,\mathbb{R})$ gauge group}\label{SL3}
This gauge group can be gauged by coupling five vector multiplets to
$N=3$ supergravity. To identify the structure constants
$f_{\Lambda\Sigma}^{\phantom{\Lambda\Sigma}\Gamma}=g\tilde{f}_{\Lambda\Sigma}^{\phantom{\Lambda\Sigma}\Gamma}$,
we define the following $SL(3,\mathbb{R})$ generators
\begin{eqnarray}
T_\Lambda=(i\lambda_2,i\lambda_7,i\lambda_5,\lambda_1,\lambda_3,\lambda_4,\lambda_6,\lambda_8)
\end{eqnarray}
where $\lambda_i$ are Gell-mann matrices. The structure constants
can be extracted from the $SL(3,\mathbb{R})$ algebra
\begin{equation}
\left[T_{\Lambda},T_\Sigma\right]=\tilde{f}_{\Lambda\Sigma}^{\phantom{\Lambda\Sigma}\Gamma}T_\Gamma\,
.
\end{equation}
\indent There are $30$ scalars transforming as
$(\mathbf{3},\bar{\mathbf{5}})+(\bar{\mathbf{3}},\mathbf{5})$ under
the $SU(3)\times SU(5)$ local symmetry. The $SO(3)$ maximal compact
subgroup of $SL(3,\mathbb{R})$ is embedded by $\mathbf{3}\rightarrow
\mathbf{3}$ and $\mathbf{8}\rightarrow \mathbf{3}+\mathbf{5}$. The $30$
scalars transform under this $SO(3)$ as
\begin{equation}
(\mathbf{3}\times\mathbf{5})+(\mathbf{3}\times\mathbf{5})=(\mathbf{3}+\mathbf{5}+\mathbf{7})+
(\mathbf{3}+\mathbf{5}+\mathbf{7}).
\end{equation}
There are accordingly no singlets under $SO(3)$ symmetry. We then
consider scalars which are singlets under $SO(2)\subset SO(3)$.
Further decomposing the above representations give six singlets,
each of these representations giving one singlet, corresponding to
the following non-compact generators of $SU(3,5)$
\begin{eqnarray}
Y_1&=&\hat{Y}_{24}+\hat{Y}_{33},\qquad
Y_2=\hat{Y}_{23}-\hat{Y}_{34},\qquad
Y_3=\hat{Y}_{15},\nonumber \\
Y_4&=&\tilde{Y}_{24}+\tilde{Y}_{33},\qquad
Y_5=\tilde{Y}_{23}-\tilde{Y}_{34},\qquad Y_6=\tilde{Y}_{15}\, .
\end{eqnarray}
With the coset representative
\begin{equation}
L=e^{\Phi_1Y_1}e^{\Phi_2Y_2}e^{\Phi_3Y_3}e^{\Phi_4Y_4}e^{\Phi_5Y_5}e^{\Phi_6Y_6},
\end{equation}
we find the following potential
\begin{eqnarray}
V&=&-\frac{1}{32}e^{-4\Phi_2-4\Phi_3}g^2\left[16\sqrt{3}e^{2\Phi_2}(e^{4\Phi_2}-1)(e^{4\Phi_3}-1)
\cosh(2\Phi_4)\cosh(2\Phi_5)\cosh(2\Phi_6)\right.\nonumber \\
&
&+\cosh^2(2\Phi_5)\left[3e^{2\Phi_3}(2e^{4\Phi_2}-3e^{8\Phi_2}-3)-12e^{2\Phi_3}(e^{4\Phi_2})^2
\cosh(4\Phi_4)\right.\nonumber \\
&
&\left.+(1+e^{4\Phi_3})\left[2(3+e^{4\Phi_2}+3e^{8\Phi_2})+(9-2e^{4\Phi_2}+9e^{8\Phi_2})
\cosh(4\Phi_4)\right]\cosh(2\Phi_6)\right]\nonumber \\
&
&+(1+e^{4\Phi_3})\cosh(2\Phi_6)\left[3+4e^{4\Phi_2}+3e^{8\Phi_2}+(3-4e^{4\Phi_2}+3e^{8\Phi_2})
\sinh^2(2\Phi_5)\right]\nonumber \\
&
&-e^{2\Phi_3}\left[3+14e^{4\Phi_2}+3e^{8\Phi_2}+3(1-6e^{4\Phi_2}+e^{8\Phi_2})\sinh^2(2\Phi_5)\right.\nonumber
\\
&
&\left.\left.-8\sqrt{3}e^{2\Phi_2}(1+e^{4\Phi_2})\cosh(2\Phi_4)\sinh(4\Phi_5)\sinh(2\Phi_6)\right]\right].
\end{eqnarray}
Apart from the trivial critical point at all $\Phi_i=0$, we have not
found any other critical points. At the trivial $AdS_4$ point, we
find
\begin{equation}
V_0=-\frac{3}{2}g^2,\qquad L^2=\frac{1}{g^2}
\end{equation}
and the scalar masses given in table \ref{table4}. Apart from the Goldstone bosons, there are marginal deformations corresponding to the scalar fields in the $\mathbf{7}$ representation of the unbroken $SO(3)$ symmetry.
\begin{table}[h]
\centering
\begin{tabular}{|c|c|c|}
  \hline
  $SO(3)$ representations & $m^2L^2\phantom{\frac{1}{2}}$ & $\Delta$  \\ \hline
  $\mathbf{3}$ & $10_{(\times 3)}$, $-2_{(\times 3)}$ & $5$, $(1,2)$  \\
  $\mathbf{5}$ & $0_{(\times 5)}$, $-2_{(\times 5)}$  & $3$, $(1,2)$ \\
  $\mathbf{7}$ & $0_{(\times 7)}$, $-2_{(\times 7)}$  & $3$, $(1,2)$ \\
  \hline
\end{tabular}
\caption{Scalar masses at the $N=3$ supersymmetric $AdS_4$ critical
point with $SO(3)$ symmetry and the corresponding dimensions of the
dual operators in $SL(3,\mathbb{R})$ gauge group}\label{table4}
\end{table}
\\
\indent We will not give the full BPS equations here due to their
complexity. To find some supersymmetric deformations of the $N=3$ SCFT dual to the $AdS_4$ critical point, we will consider a truncation to $\Phi_1$, $\Phi_2$ and $\Phi_3$. Within this truncation, we find that $S_{AB}=\frac{1}{2}W\delta_{AB}$ and the system of BPS equations
\begin{eqnarray}
\Phi_1'&=&0,\\
\Phi_2'&=&\pm\frac{1}{2}\frac{\pd W}{\pd\Phi_2}=\mp \sqrt{3}g\cosh(2\Phi_2)\sinh(\Phi_3),\\
\Phi_3'&=&\pm\frac{\pd W}{\pd\Phi_3}=\mp g\left[\sqrt{3}\cosh\Phi_3\sinh(2\Phi_2)+\sinh\Phi_3\right],\\
A'&=&\mp W
\end{eqnarray}
where the superpotential is given by
\begin{equation}
W=-g\left[\cosh\Phi_3+\sqrt{3}\sinh(2\Phi_2)\sinh\Phi_3\right].
\end{equation}
With the scalar kinetic terms
\begin{equation}
-\frac{1}{4}e^{-4\Phi_2}(1+e^{4\Phi_2})^2\Phi_1'^2-\Phi_2'^2-\frac{1}{2}\Phi_3'^2,
\end{equation}
the scalar potential can be written as
\begin{eqnarray}
V&=&-\frac{1}{4}\frac{\pd W}{\pd \Phi_2}-\frac{1}{2}\frac{\pd W}{\pd \Phi_3}-\frac{3}{2}W^2\nonumber \\
&=&-\frac{1}{4}g^2\left[2+\cosh(2\Phi_3)+\cosh(4\Phi_2-)[9\cosh(2\Phi_3)-6]\right.\nonumber \\
& &\left.+8\sqrt{3}\sinh(2\Phi_2)\sinh(2\Phi_3)
\right].
\end{eqnarray}
We now analyze asymptotic behaviors of the solution
near the UV and IR of the flow. Near the $AdS_4$ critical point, we
find
\begin{equation}
\frac{1}{\sqrt{3}}\Phi_2+\Phi_3\sim e^{-3g_1r}\sim e^{-\frac{3r}{L}},\qquad \Phi_3-\frac{\sqrt{3}}{2}\Phi_2\sim e^{2g_1r}\sim e^{\frac{2r}{L}},\qquad A\sim g_1r\sim\frac{r}{L}\, .
\end{equation}
We see that $\frac{1}{\sqrt{3}}\Phi_2+\Phi_3$ is dual to a vacuum
expectation value of a marginal operator while
$\Phi_3-\frac{\sqrt{3}}{2}\Phi_2$ is dual to an irrelevant operator
of dimension $\Delta=5$. Since a marginal operator does not break
conformal symmetry, we expect that the flow involves the operator
dual to $\Phi_3-\frac{\sqrt{3}}{2}\Phi_2$. In this case, the UV SCFT
dual to the supersymmetric $AdS_4$ critical point should appear in
the IR since the operator driving the flow is irrelevant at the
fixed point.
\\
\indent Near the singularity, we find for large $|\Phi_2|$,
\begin{eqnarray}
& &\Phi_3\sim\Phi_2\sim\mp\frac{1}{3}\ln \left[\frac{3\sqrt{3}gr}{4}\right],\qquad A\sim \frac{1}{3}\ln r,\nonumber \\
& &ds^2=r^{\frac{2}{3}}dx^2_{1,2}+dr^2\, .
\end{eqnarray}
This leads to a physical singularity and describes an RG flow in the
dual $N=3$ supersymmetric field theory to a conformal fixed point in the
IR.

\section{$SO(2,1)\times SO(2,2)$ gauge group}\label{SO2_2}
The last gauge group to be considered in this paper is $SO(2,1)\times
SO(2,2)\sim SO(2,1)\times SO(2,1)\times SO(2,1)$. This gauge group
can be obtained by coupling six vector multiplets to $N=3$
supergravity with the following structure constants
\begin{equation}
f_{\Lambda\Sigma}^{\phantom{\Lambda\Sigma}\Gamma}=(g_1\epsilon_{\bar{A}\bar{B}\bar{D}}\eta^{\bar{D}\bar{C}},
g_2\epsilon_{\bar{i}\bar{j}\bar{l}}\eta^{\bar{l}\bar{k}},g_3\epsilon_{\tilde{i}\tilde{j}\tilde{l}}
\eta^{\tilde{l}\tilde{k}})
\end{equation}
where $\bar{A},\bar{B},\ldots =1,4,5,$,
$\bar{i},\bar{j},\ldots=2,6,7$, $\tilde{i},\tilde{j},\ldots=3,8,9$
and
\begin{equation}
\eta_{\bar{A}\bar{B}}=\textrm{diag}(1,-1,-1),\qquad
\eta_{\bar{i}\bar{j}}=\textrm{diag}(1,-1,-1),\qquad
\eta_{\tilde{i}\tilde{j}}=\textrm{diag}(1,-1,-1)\, .
\end{equation}
\indent At the vacua, the full gauge group $SO(2,1)\times SO(2,2)$
will be broken to its maximal compact subgroup $SO(2)\times
SO(2)\times SO(2)$. We will consider scalars which are invariant
under the $SO(2)\times SO(2)$ residual symmetry chosen to be the first two $SO(2)$'s. In this case, there
are twelve singlets given by
\begin{eqnarray}
Y_1&=&\hat{Y}_{15},\qquad Y_2=\hat{Y}_{16},\qquad
Y_3=\hat{Y}_{25},\qquad Y_4=\hat{Y}_{26},\nonumber \\
Y_5&=&\hat{Y}_{35},\qquad Y_6=\hat{Y}_{36},\qquad
Y_7=\tilde{Y}_{15},\qquad Y_8=\tilde{Y}_{16},\nonumber \\
Y_9&=&\tilde{Y}_{25},\qquad Y_{10}=\tilde{Y}_{26},\qquad
Y_{11}=\tilde{Y}_{35},\qquad Y_{12}=\tilde{Y}_{36}\, .
\end{eqnarray}
The coset representative can be parametrized by
\begin{equation}
L=\prod_{i=1}^{12}e^{\Phi_iY_i}\, .
\end{equation}
The potential is highly complicated. We refrain from giving its
explicit form here but only note that the resulting potential admits
a Minkowski vacuum at $\Phi_i=0$, for $i=1,\ldots ,12$ preserving
$N=3$ supersymmetry and $SO(2)\times SO(2)\times SO(2)$ symmetry. It
can also be checked that there are precisely six massless Goldstone
bosons of the symmetry breaking $SO(2,1)\times SO(2,2)\rightarrow
SO(2)\times SO(2)\times SO(2)$.

\section{Conclusions}\label{conclusions}
In this paper, we have studied $N=3$ gauged supergravity in four
dimensions with various types of semisimple gauge groups and
classified their vacua. We now summarize the main results found in
this paper. For $SO(3)\times SO(3)$, $SO(3,1)$ and
$SL(3,\mathbb{R})$ gauge groups, there exists a maximally
supersymmetric $AdS_4$ critical point at which all scalars
vanishing. The critical point has $SO(3)$ symmetry in $SO(3,1)$ and
$SL(3,\mathbb{R})$ gauge groups and $SO(3)\times SO(3)$ symmetry for
$SO(3)\times SO(3)$ gauge group. In the latter case, we have also found a
non-trivial $AdS_4$ critical point with $SO(3)_{\textrm{diag}}$
symmetry and unbroken $N=3$ supersymmetry. A
holographic RG flow interpolating between the $SO(3)\times SO(3)$
and $SO(3)_{\textrm{diag}}$ critical points including a number of RG flows to
non-conformal gauge theories have also been given. The non-conformal RG flows break conformal symmetry but preserve $N=3$ or $N=1$ supersymmetries. A similar study has also been carried out in the case of $SO(3,1)$ and $SL(3,\mathbb{R})$. These results
might be useful in the holographic study of $N=3$
Chern-Simons-Matter theories in three dimensions.
\\
\indent For $SO(2,1)\times SO(2,2)$ gauge group, the gauged
supergravity admits $N=3$ Minkowki vacuum when all scalars vanish.
In the case of $SO(2,1)$ and $SO(2,2)\sim SO(2,1)\times SO(2,1)$
gauge groups, the resulting gauged supergravities admit a
half-maximal supersymmetric domain wall as a supersymmetric vacuum.
This solution should be useful in the context of the DW/QFT
correspondence for studying strongly coupled gauge theories in three
dimensions. When all scalars vanish, there exists a
non-supersymmetric $AdS_4$ critical point with $SO(2)$ and
$SO(2)\times SO(2)$ symmetries, respectively. This critical point
and all of the non-supersymmetric critical points identified in this
paper are unstable.
\\
\indent The results have some similarity to the gaugings of $N=2$ gauged supergravity in seven dimensions studied in \cite{Non_compact_7DN2}, but in the present case, non-conformal flows with partially broken supersymmetry are possible. It should also be remarked that although we have
considered only the $G_0$ part of the full semisimple gauge group
$G_0\times H$, all of the vacua and solutions we have found are
valid in the full theory with $G_0\times H$ gauge group. This is a
consequence of the fact that all scalars in $SU(3,n)/SU(3)\times SU(n)\times U(1)$ we have considered are $H$
singlets. By the argument given in \cite{warner}, solutions
identified within the scalar submanifold parametized by $H'\times H$, with $H'\subset G_0$,
singlets are solutions of the full $G_0\times H$ theory.
\\
\indent It would be interesting to identify the SCFTs or
non-conformal gauge theories that are dual to the gravity solutions
obtained here. Looking for more general domain walls with the truncated scalars restored could be useful since these scalars correspond to relevant operators. From the analysis of this paper, we expect these scalars to break some supersymmetry. Investigating their role in the dual SCFT should give some insight to relevant deformations of the dual three-dimensional SCFT. It would also be interesting to look for supersymmetric Janus solutions which are dual to some conformal interface in the $N=3$ SCFTs. A number of these solutions have been obtained within the maximal $N=8$ gauged supergravity in \cite{warner_Janus}.
\\
\indent
It should be noted that all gaugings considered here are of
``electric'' type in which only electric gauge fields are involved.
Similar to the maximal $N=8$ and half-maximal $N=4$ gauged
supergravities \cite{N8_embedding_gauged,N4_embedding_gauge}, it
could be interesting to apply the embedding tensor formalism to
the $N=3$ gauged supergravity and look for more general gaugings
such as the magnetic or dyonic gaugings in which magnetic gauge
fields also participate in the process of gauging. We then expect
many other possible gauge groups will arise from the embedding
tensor formalism similar to the $N=4$ gauged supergravity with
$SU(1,1)/U(1)\times SO(6,n)/SO(6)\times SO(n)$ scalar manifold
studied in \cite{N4_embedding_gauge}.
\\
\indent The $N=3$ $AdS_4$ critical point with $SO(3)\times SO(3)$
symmetry within the dyonic $ISO(7)$ gauged $N=8$ supergravity is
known \cite{Henning_4D_vacua,dyonic_ISO7,N3_ISO7}, and the holographic study of the possible dual
SCFT has been given in \cite{holographic_N3_ISO7_Guarino,holographic_N3_ISO7}. Furthermore, this
$N=3$ $AdS_4$ solution has known massive type IIA origin \cite{holographic_N3_ISO7_Guarino,ISO7_from_IIA_S6}.
Similarly, investigating the embedding of the results presented in
this paper in higher dimensions could be of interest and will give
rise to new $N=3$ $AdS_4$ backgrounds within the context of
string/M-theory. We will leave these interesting issues for future
works.
\vspace{0.5cm}\\
{\large{\textbf{Acknowledgement}}} \\
This work is supported by Chulalongkorn University through
Ratchadapisek Sompoch Endowment Fund under grant GF-58-08-23-01 (Sci-Super II).


\end{document}